%% file: main.tex
\documentclass[runningheads]{llncs}
\usepackage{graphicx}
\usepackage{tikz}
\usepackage{amsmath}

\usepackage{filecontents}

\usepackage{amssymb}
\usepackage{array}
\usepackage{booktabs}
\usepackage{fullpage}
\usepackage{threeparttable}
\usepackage{wasysym}
\usepackage{flushend}
\usepackage{caption}
\usepackage{subcaption}
\usepackage{longtable}
\usepackage{scrtime}
\usepackage[multiple]{footmisc}
\usepackage{footnote}
\usepackage{makecell}
\usepackage{multirow}
\usepackage{rotating}
\usepackage{subcaption}
\usepackage{multicol}
\usepackage{graphicx}
\usepackage{float}
\usepackage{listings}
\usepackage{textcomp}
\usepackage{footnote}
\usepackage{enumitem}
\usepackage{lipsum}

\usepackage{epsfig,hyperref,color,graphicx,xspace,pifont,dirtree}
\usepackage{cleveref}

\usetikzlibrary{shapes.geometric, arrows}
\usepackage{pgfplotstable}

\usepackage{pgfplots}
    \pgfplotsset{compat=1.3}
    
    \definecolor{bblue}{HTML}{4F81BD}
    \definecolor{rred}{HTML}{C0504D}
    \definecolor{ggreen}{HTML}{9BBB59}

\newcommand{\rom}[1]{{\em\lowercase\expandafter{(\romannumeral #1\relax)}}}

\newcommand\cve[1]{{\href{https://cve.mitre.org/cgi-bin/cvename.cgi?name=CVE-#1}{CVE-#1}}}

\lstdefinestyle{customc}{
  belowcaptionskip=1\baselineskip,
  breaklines=true,
  frame=ltrb,
  xleftmargin=\parindent,
  language=C,
  showstringspaces=false,
  basicstyle=\footnotesize\ttfamily\bfseries,
  keywordstyle=\bfseries\color{green!60!black},
  commentstyle=\itshape\color{purple!60!black},
  identifierstyle=\color{blue},
  stringstyle=\color{red},
}
\lstdefinestyle{customasm}{
  belowcaptionskip=1\baselineskip,
  frame=R,
  xleftmargin=\parindent,
  language=[x86masm]Assembler,
  basicstyle=\footnotesize\ttfamily\bfseries,
  commentstyle=\itshape\color{purple!60!black},
}

\newcommand{\tool} {{ MicroGuards}\xspace}
\newcommand{\stool} {{ MicroGuard}\xspace}
\newcommand{\tooltitle} {{\sc \textbf{MicroGuards}}\xspace}

\newcommand\chk{\ding{51}}
\begin{document}
\title{\Large \bf Enabling Lightweight Privilege Separation in Applications with~\tooltitle \thanks{This paper will appear at the ACNS-SecMT2023 (Security in Mobile Technologies).}}
\author{Zahra Tarkhani\inst{1}\thanks{This work was done when the author was affiliated with the University of Cambridge.}\and
Anil Madhavapeddy\inst{2} }
%

\institute{Microsoft Research Cambridge\\
 \and
University of Cambridge\\
}
\maketitle              % typeset the header of the contribution
\begin{abstract}
Application compartmentalization and privilege separation are our primary weapons against ever-increasing security threats and privacy concerns on connected devices. Despite significant progress, it is still challenging to privilege separate inside an application address space and in multithreaded environments, particularly on resource-constrained and mobile devices. We propose MicroGuards, a lightweight kernel modification and set of security primitives and APIs aimed at flexible and fine-grained in-process memory protection and privilege separation in multithreaded applications. MicroGuards take advantage of hardware support in modern CPUs and are
high-level enough to be adaptable to various architectures. This paper focuses on enabling MicroGuards on embedded and mobile devices running Linux kernel and utilizes tagged memory support to achieve good performance. Our evaluation show that MicroGuards add small runtime overhead (less than 3.5\%), minimal memory footprint, and are practical to get integrated with existing applications to enable fine-grained privilege separation.

%\keywords{security, privilege separation}
\end{abstract}
\section{Introduction}

More than ever, we depend on highly connected computing systems in today's world, where over 6.3 Billion people use smartphones, and 35.82 billion IoT (Internet of Things) devices are installed worldwide~\cite{iotall}. Our growing reliance on edge-cloud services in recent years has been constantly and increasingly threatened by a wide range of security and privacy breaches at scales never seen before\cite{breach1,breach2,breach3,tarkhani2022enhancing}. The attack surface of modern applications includes a mixture of traditional attack vectors with new threats within/across various dependencies and system abstractions. 

Many software attacks target sensitive content in an application's address space, usually through remote exploits, malicious third-party libraries, or unsafe language vulnerabilities.  
Processing highly sensitive data in a single large compartment (e.g.,~process or enclave) leads to real threats that require effective protection against:~\rom{1} attackers can exploit vulnerabilities in less secure parts of the code to leak information, escalate privileges, or take control of the application or even the host.
~\rom{2} an application's secret data (e.g.,~private keys or user passwords) can be leaked in the presence of untrusted code parts or compromised third-party libraries like OpenSSL~\cite{durumeric2014matter};
~\rom{3} privileged functions or modules can be misused to access private content~\cite{deng2015iris};
~\rom{4} applications written in memory-safe languages such as Rust or OCaml are vulnerable via unsafe external libraries that jeopardize all other safety guarantees~\cite{almohri2018fidelius,lamowski2017sandcrust};
and
~\rom{5} in multithreaded use cases, attackers can exploit vulnerabilities (e.g.,~TOCTOU or buffer overflows) so the compromised thread can access sensitive data owned by other threads~\cite{tserver}.
This whole class of attacks could be avoided by providing a practical way to enforce the least privilege within a shared address space. Table~\ref{t:cve-table} summarizes some of these real threats that intra-process protection is effective against.

Hence, the importance of in-address space security threats results in significant improvement in hardware support for efficient memory isolation~\cite{mte,mpk,arm2012architecture,watson2015cheri}. 
However, existing simple APIs for utilizing such hardware features are not effective due to the complexity of attacks as well as various hardware limitations~\cite{vahldiek2019erim,park2018libmpk,chen2016shreds} in security and performance particularly for resource constrained devices. 
These systems mainly require specific programming languages or rely on x86 features which are not practical for wide range of IoT and mobile devices.

\input{cve_tab.tex}

Many security-sensitive applications such as OpenSSH~\cite{provos2003preventing} rely on process-based isolation to separate their components into different privileged processes. However, this usually requires redesigning an application from scratch using a multiprocess architecture (e.g., Chrome) and is difficult for many multithreaded applications such as web servers. 
Previous work such as Privtrans~\cite{brumley2004privtrans} and Wedge~\cite{bittau2008wedge} provide automatic process-based isolation of applications with a huge overhead ($\approx 80\%-40x$ slowdown). \\
Conventional process abstractions such as \texttt{fork} introduce security and efficiency issues~\cite{baumann2019fork}, and alternatives such as \texttt{clone} are not fine-grained enough to switch between data sharing and copying between process address spaces for security-critical resources. This lack of flexibility in the underlying interfaces means developers cannot easily prevent in-process attacks, and so multithreaded applications are difficult to privilege separate. This class of attacks could be avoided by providing a practical way to protect memory within an address space.

\begin{figure}[t]
\centering
\includegraphics[scale=.6]{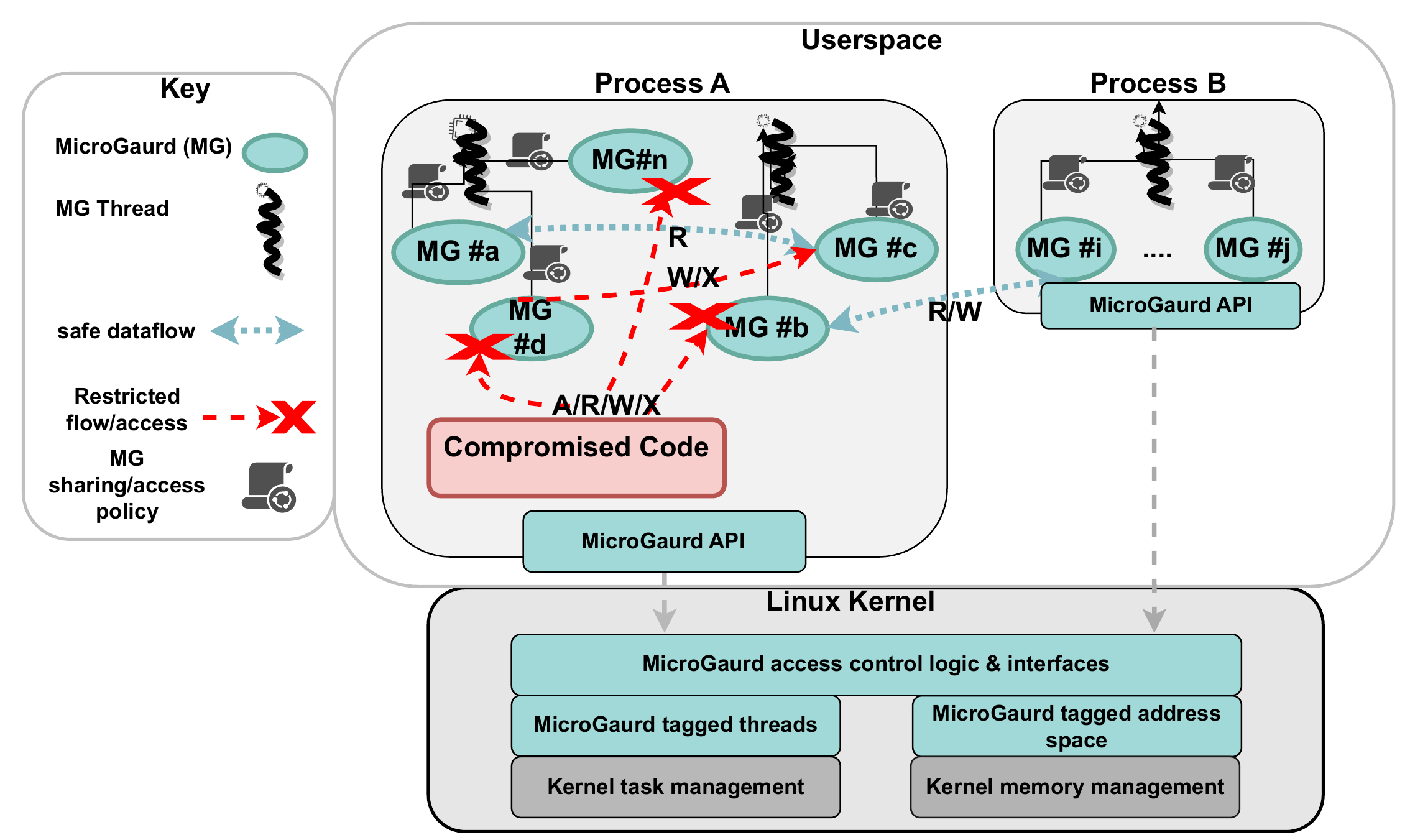}

\caption{High-level architecture of~\tool: it provides in-process isolation as well as thread-granularity privilege separation so each~\stool thread can tag itself, its address space, and define its own trust boundaries.}
\label{hlevel}
\end{figure}

\input{compare.tex}

In this paper, we present~\tool, a new OS abstraction for enforcing least privilege on slabs of memory within the same address space. It takes advantage of modern hardware features to provide a flexible and efficient way to define trust boundaries to isolate sensitive data while supporting familiar APIs for secure multithreading and memory management. We provide a virtual memory tagging and access control abstraction within the kernel, then extend the kernel to support mapping~\tool to threads; hence, any thread can selectively protect or share its memory compartments from untrusted code within itself or from any untrusted thread (see Figure~\ref{hlevel}). 

 Hence, we designed a new memory compartmentalisation abstraction to overcome this limitation efficiently. 
\tool virtual memory tagging layer bypasses most of the kernel's paging abstraction to enable isolated blocks of tagged memory which could be mapped to the undelying hardware features such as ARM MD (memory domains) or MTE (memory tagged extension) for stronger isolation enforcement and performance optimization. 
Moreover, these hardware features are difficult to use securely (require a strong access control mechanism) and portably due to differing semantics across the Linux Kernel virtual memory abstraction and hardware provided features (\S\ref{back}). Note that~\tool virtual memory layer can also be enabled with available simple address space translation mechanism and without hardware-based memory tagging capabilities. However, it is specifically designed for properly utilizing such beneficial hardware security features. Hence,~\tool is a high-level OS abstraction that aims to:

\begin{itemize}
\item develop a new kernel-assisted mechanism based on mutual-distrust for intra-process privilege separation that supports isolating private contents, a secure multithreading model, and secure communication within a shared address space. 
\item explain how to utilize modern CPU facilities for efficient memory tagging to avoid the overhead of existing solutions (due to TLB flushes, per-thread page tables, or nested page table management).
\item show that the implementation is sufficiently lightweight ($\approx 5K$ LoC) to be practical for IoT and mobile devices with a minimal memory footprint.
\item evaluate our implementation using real-world software such as Apache HTTP server, OpenSSL, and Google's LevelDB, which shows  
\tool add negligible runtime overhead for lightly modified applications.
\end{itemize}

The remainder of this paper elaborates on the CPU hardware features we use (\S\ref{back}), describes the architecture (\S\ref{over}) and implementation of~\tool (\S\ref{imp}), presents an evaluation (\S\ref{eval}) and the tradeoffs of our approach (\S\ref{diss}).

\section{Background}\label{back}

\subsection{ARM VMSA}\label{sec:vmsa}

ARM virtual memory system architecture (VMSA) is tightly integrated with the security extensions, the multiprocessing extensions, the Large Physical Address Extension (LPAE), and the virtualization extensions. VMSA provides MMUs that control address translation, access permissions, and memory attribute determination and checking for memory accesses. The extended VMSAv7/v8 provides multiple stages of memory system control; for operation in Secure state (e.g.,~EL1\&0 stage 1 MMU) and for operation in Non-secure state (e.g.,~EL2 stage 1 MMU, EL1\&0 stage 1 MMU, and EL1\&0 stage 2 MMU). VMSAv8.5 adds more MMUs for additional isolation in the secure world.
Each MMU uses a set of address translations and associated memory properties held in TLBs. If an implementation does not include the security extensions, it has only a single security state, with a single MMU with controls equivalent to the Secure state MMU controls. A similar argument is valid for when am implementation does not include the virtualization extensions.

System Control coprocessor (CP15) registers control the VMSA, including defining the location of the translation tables. They include registers that contain memory fault status and address information. 
The MMU supports memory accesses based on memory sections or pages, supersections consist of 16MB blocks of memory, sections consist of 1MB blocks of memory or 64KB blocks of memory, and pages consist of 4KB blocks of memory. Operation of MMUs can be split between two sets of translation tables, defined by the Secure and Non-secure copies of \texttt{TTBR0} and \texttt{TTBR1}, and controlled by \texttt{TTBCR}. For \texttt{hyp} mode stage 1, The \texttt{HTTBR} defines the translation table for EL2 MMU, controlled by \texttt{HTCR}. For stage 2 translation, The \texttt{VTTBR} defines the translation table, controlled by \texttt{VTCR}. Access to a memory region is controlled by the access permission bits and the domain field in the TLB entry.

\subsubsection{ARM memory domains (MDs)}\label{memdomain}
A domain is a collection of contiguous memory regions. The ARM VMSAv7 architecture supports 16 domains, and each VMSA memory region is assigned to a domain. First-level translation table entries for page tables and sections include a domain field. Translation table entries for super-sections do not include a domain field (super-sections are defined as being in domain 0). Second-level translation table entries inherit a domain setting from the parent first-level page table entry. Each TLB entry includes a domain field. A domain field specifies which domain the entry is in, and a two-bit field controls access to each domain in the Domain Access Control Register (\texttt{DACR}). Each field enables access to an entire domain to be enabled and disabled very quickly without TLB flushes so that whole memory areas can be swapped in and out of virtual memory very efficiently. Hence \texttt{DACR} controls the behavior of each domain and is not guarded by the access permissions for TLB entries in that domain. Also, \texttt{DACR} defines the access permission for each of the sixteen isolation domains. The \texttt{DACR} is a 32-bit read/write register and is accessible only in privileged modes. When the security extensions are implemented \texttt{DACR} is a banked register, and write access to the secure copy of the register is disabled when the \texttt{CP15SDISABLE} signal is asserted high.
To access the \texttt{DACR} you read or write the \texttt{CP15} registers. For example:
'\texttt{MRC} $p15,0,<Rt>,c3,c0,0$ ' for reading from \texttt{DACR}  and '\texttt{MCR} $p15,0,<Rt>,c3,c0,0$' for writtig to \texttt{DACR}.
Data Fault Status Register (DFSR) holds status information about the last data fault in MDs. It is a 32-bit read/write register, accessible only in privileged modes. These registers are banked when security extensions are enabled, so we could have separate 16 domains inside TrustZone secure world as well as the normal world.

\input{domains.tex}

The four possible access rights for a domain are No Access, Manager, Client, and Reserved (see Table~\ref{domains}).  Those fields let the processor ~\rom{1} prohibit access to the domain mapped memory--No Access;~\rom{2} allow unlimited access to the memory despite permission bits in the page table-- Manager; or~\rom{3} let the access right be the same as the page table permissions--Client. Any access violation causes a domain fault, and changes to the DACR are low cost and activated without affecting the TLB. 

ARM MDs look like a good building block for in-process memory protection.  Changing domain permissions does not require TLB flushes, and they do not require extensive modifications to the kernel memory management structures that might otherwise introduce security holes due to inevitable TLB and memory management bugs~\cite{tlbbug}.

Though ARM MDs are a useful isolation primitive in concept, the current hardware implementation and OS support suffer from significant problems that have prevented their broader adoption:\\
\noindent\textbf{Scalability:} ARM relies on a 32-bit \texttt{DACR} register and so supports only up to 16 domains. Allocating a larger register (e.g.,~512 bits) would mean larger page table entries or additional storage for domain IDs.\\
\textbf{Flexibility:}
Unlike Intel MPK, ARM-MDs only apply to first-level entries; the second-level entries inherit the same permissions. This prevents arbitrary granularity of memory protections to small page boundaries and reduces the performance of some applications~\cite{cox2017efficient}.
Also, the \texttt{DACR} access control options do not directly mark a domain as read-only, write-only, or exec-only. So the higher-level VM abstraction should resolve these issues.\\
\textbf{Performance:} Changing the \texttt{DACR} is a fast but privileged operation, so any change of domain access permissions from userspace require a system call. This is unlike Intel MPK that makes its Protection Key Rights Register (PKRU) accessible directly from userspace.\\
\textbf{Userspace:} There is no Linux userspace interface for using ARM-MD; it is only used within the kernel to map the kernel and userspace into separate domains. In contrast, Linux already provides some basic support for utilizing Intel MPK from userspace.\\
\textbf{Security:} Though the \texttt{DACR} is only accessible in privileged mode, any syscall that changes this register is a potential breach that could cause the attacker to gain full control of the host kernel (e.g.,~through the misuse of the \texttt{put\_user/get\_user} kernel API in~\cve{2013-6282}). Also, since only 16 domains are supported, guessing other domains' identifiers is trivial, making it essential not to expose these directly to application code.

\subsubsection{Address space identifier}

The VMSA permits TLBs to hold any translation table entry that does not directly cause a translation fault or an access flag fault.
To reduce the software overhead of TLB maintenance, the VMSA differentiates between \textit{global pages} and \textit{process-specific pages} through the Address Space Identifier (ASID). A global virtual memory page is available for all processes on the system, and a single cache entry can exist for this page translation in the TLB. A non-global virtual memory page is process-specific, associated with a specific ASID. 
The ASID identifies pages associated with a specific process and provides a mechanism for changing process-specific tables without maintaining the TLB structures. Hence, multiple TLB entries can exist for the same page translation, but only TLB entries that are associated with the current ASID are available to the CPU (x86 supports a similar mechanism, called PCID). On ARMv7, the current ASID is defined by the Context ID Register (CONTEXTIDR), and on ARMv8, the ASID is defined by the translation table base registers that causes better performance compare to ARMv7. 
Each \texttt{TTBR} contains an ASID field, and the \texttt{TTBCR.A1} field selects which ASID to use. If the implementation supports 16 bits of ASID, then the upper 8 bits of the ASID must be written to 0 by software when the context being affected only uses 8 bits. ASIDs/PCIDs are useful for relatively faster context switching~\cite{litton2016light} and more efficient page table isolation as shown in design of kernel page-table isolation (KPTI or PTI, previously called KAISER~\cite{gruss2017kaslr}) for mitigating Meltdown vulnerability~\cite{lipp2018meltdown}. 

\subsubsection{MTE \& PAC}

Memory Tagging Extension (MTE), also called memory coloring, is introduced in Armv8.5-A.
Memory locations are tagged by adding four bits of metadata to each 16 bytes of physical memory (this is the Tag Granule). Tagging memory implements the lock. Hence, pointers and virtual addresses are modified to contain the key. In order to implement the key bits without requiring larger pointers, MTE uses the TBI (top byte ignore) feature of the Armv8-A Architecture. When TBI is enabled, the top byte of a virtual address is ignored when using it as an input for address translation similar to PAC (Pointer Authentication Code) design. This allows the top byte to store metadata. Memory tagging and pointer authentication both use the upper bits of an address to store additional information about the pointer: a tag for memory tagging, and a PAC for pointer authentication.
Both technologies can be enabled at the same time. The size of the PAC is variable, depending on the size of the virtual address space. When memory tagging is enabled at the same time, there are fewer bits available for the PAC.

MTE adds a new memory type, Normal Tagged Memory, to the Arm Architecture. 
A mismatch between the tag in the address and the tag in memory can be configured to cause a synchronous exception or to be asynchronously reported. When the asynchronous mode is enabled, upon fault, the PE updates the \texttt{TFSR\_EL1} register. Then the kernel detects the change during context switching, return to \texttt{EL0}, kernel entry from \texttt{EL1}, or kernel exit to \texttt{EL1}. 
MTE is currently supported by LLVM, and when it is enabled, a call to \texttt{malloc()} will allocate the memory and assign a tag for the buffer. The returned pointer will include the allocated tag. If software using the pointer goes beyond the limits of the buffer, the tag comparison check will fail. This failure will allow us to detect the overrun.
Similarly, for use-after-free, on the call to \texttt{malloc()} the buffer gets allocated in memory and assigned a tag value. The pointer that is returned by \texttt{malloc()} includes this tag. The C library might change the tag when the memory is released. If the software continues to use the old pointer, it will have the old tag value, and the tag-checking process will catch it.

\begin{figure*}
\centering
\includegraphics[scale=.5]{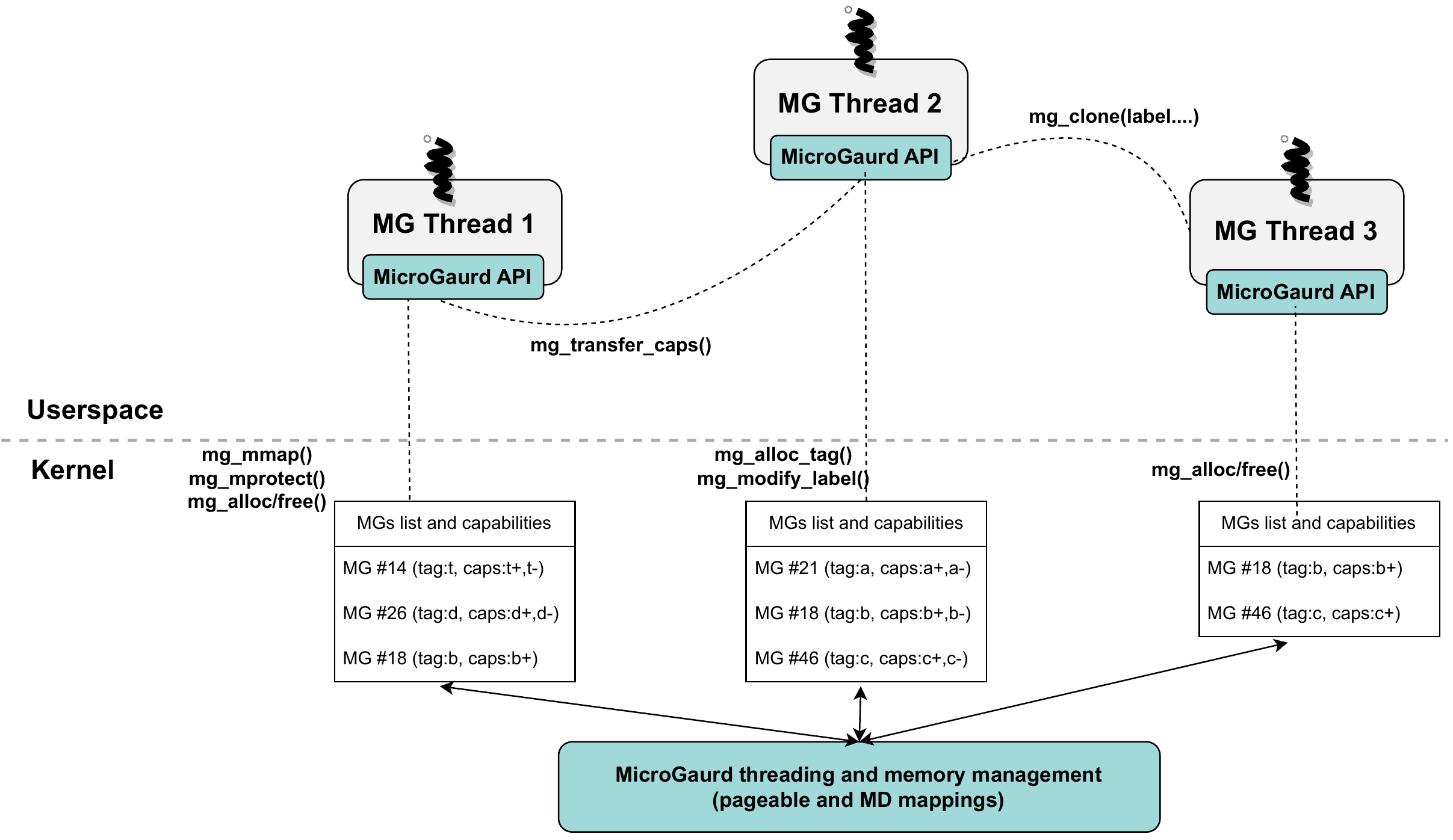}
\caption{simple~\tool simple threading example: each~\stool thread is a security principal, it can define security policies for controlling its own~\tool collection, and pass its capabilities to other threads for secure sharing. The kernel then enforces~\stool security policies and handles its virtual memory management.}
\label{threading}
\end{figure*}

\section{\tool}\label{over}

We now describe the implementation of~\tool, which is an abstraction over the underlying kernel and hardware memory management for efficient intra-process isolation.~\tool abstraction has an emphasis on security, performance, and extensibility to support various hardware memory tagging primitives through a higher-level interface that hides the hardware limitations (\S\ref{memdomain}).

\subsection{Design Principles}\label{dp}

The~\tool interface aims to enforce least privilege principle for memory accesses via the following guidelines:\\
\textbf{Fine-grained strong isolation:}  All threads of execution should be able to define their security policies and trust models to selectively protect their sensitive resources. Current OS security models of sharing (``everything-or-nothing'') are not flexible enough for defining fine-grained trust boundaries within processes or threads (lightweight processes).\\
\textbf{Performance:} Launching~\tool, changing their access permissions, sharing across processes, and communications through capability passing should have minimal overhead. Moreover, untrusted (i.e., ~\tool-independent) parts of applications should not suffer any overhead.\\
\textbf{Efficiency:}~\tool should be lightweight enough even for mobile and IoT devices running on a few megabytes of memory and slow ARM CPUs.\\
\textbf{Compatibility:} It is difficult to provide strong security guarantees with no code modifications, and~\tool is no exception. We move most of these modifications into the Linux kernel (increasingly popular for embedded deployments~\cite{iot19}) and provide simple userspace interfaces.~\tool should be implemented without extensive changes to the Linux and not depend on a specific programming language, so existing applications can be ported easily.\\

To achieve fine-grained isolation with mutual-distrust, we need a security model that lets each thread protect its own~\tool from untrusted parts of the same thread as well as other threads and processes.
 Simply providing POSIX memory management (e.g. \texttt{malloc} or \texttt{mprotect}) is inadequate. As a simple example, attackers can misuse the API for changing the memory layout of other threads~\tool or unauthorized memory allocation. 
The~\tool interface needs to ~\rom{1} provide isolation within a single thread;~\rom{2} be flexible for sharing and using~\tool between threads, and ~\rom{3} provides the capability to restrict unauthorized permission changes or memory mappings modification of allocated~\tool. Previous work such as ERIM~\cite{vahldiek2019erim} or libMPK~\cite{park2018libmpk}
does not offer such security guarantees since their focus is more on performance and domain virtualization.

We derive inspiration from Decentralized Information Flow Control (DIFC)~\cite{krohn2007information} but with a more constrained interface -- by not supporting information flow within a program, we avoid the complexities and performance overheads that typically involves.
 Existing DIFC kernels such as HiStar~\cite{zeldovich2006making} achieve our isolation goals, but requires a non-POSIX-based OS that opposes our compatibility goal. To have a practical and lightweight solution, we therefore built~\tool over a modified Linux kernel, and internally utilizing modern hardware facilities such as ARM MDs for good performance.

\subsection{Threat Model \& Assumptions}

This paper focuses on two types of threats. First, memory-corruption based threats inside a shared address space that lead to sensitive information leakage; these threats can be caused by bugs or malicious third-party libraries (see Table~\ref{cvetab}). Second, attacks from threads that could get compromised by exploiting logical bugs or vulnerabilities (e.g., buffer overflow attacks, code injection, or ROP attacks). We assume the attacker can control a thread in a vulnerable multithreaded application, allocate memory, and fork more threads up to resource limits by the OS and hardware. The attacker will try to escalate privileges through the attacker-controlled threads or gain control of another thread, e.g., by manipulating another thread's data or via code injection. The adversary may also bypass protection regions by exploiting race conditions between threads or by leveraging confused-deputy attacks.
 
\tool thus provides isolation in two stages: firstly within a single thread (through mg\_lock/unlock calls), and then across threads in the same process. We consider threads to be security principals that can define their security policies based on mutual-distrust within the shared address space. We protect each thread's~\tool against unauthorized, accidental, and malicious access or disclosure. Therefore, the TCB consists of the OS kernel, which performs this enforcement. It also assumes developers correctly specify their policies through the userspace interface for allocating~\tool and transferring capabilities.

\tool are not protected against covert channels based on shared hardware resources (e.g., a cache). Systems such as Nickel~\cite{sigurbjarnarson2018nickel} or hardware-assisted platforms such as Hyperflow~\cite{ferraiuolo2018hyperflow} could be a helpful future addition for side-channel protection on~\tool.

\input{utiles_acc.tex}
\subsection{\tool Access Control Mechanism}\label{labeling}
Each~\stool is a contiguous allocation of memory that (by default) only its owner thread can access, add/remove pages to/from it, and change its access permission. Our modified Linux kernel enforces the access control via a dynamic security policy based on DIFC~\cite{zeldovich2006making} and a simpler version of the Flume~\cite{krohn2007information} labeling model.

 Each~\stool thread $t$ has one label ${L_{t}}$ that is the set of its unique tags. Privileges are represented in forms of two capabilities $\theta^{+}$  and $\theta^{-}$ per tag $\theta$ for adding or removing tags to/from labels. 
These capabilities are stored in a capability list $C_{p}$ per thread $p$.
To improve its performance,~\tool have only one unique secrecy tag assigned internally by the kernel when created by \texttt{mg\_create}. For improving security, none of~\tool API propagates tags in the userspace; all APIs access control is done internally within the kernel. The kernel allows information flow from $\alpha$ to $\beta$ only if $L_{\alpha}\subseteq L_{\beta}$. 
Every thread $p$ may change its label from $L_{i}$ to $L_{j}$ if it has the capability to add tags present in $L_{j}$ but not in $L_{i}$, and can drop the tags that are in $L_{i}$ but not in $L_{j}$. This is formally declared as ($L_{j}-L_{i}\subseteq C_{p}^{+}) \land (L_{i}-L_{j}\subseteq C_{p}^{-} )$.

When a thread has $\theta^{+}$ capability for~\stool $\theta$, it gains the privilege to only access~\stool $\theta$ with the permission set by its owner (read/write/execute).
The access privileges to each~\stool can be different; hence, two threads can share a~\stool, but the access privileges can differ. \\
 Having a $\theta^{-}$ capability lets it declassify~\stool $\theta$. This allows the thread to modify the~\stool memory layout by add/remove pages to it, change permissions, or copy the content to untrusted sources.Unsafe operations like declassification r equire the thread to be an owner or an authority (\texttt{acts-for} relationship) then via \texttt{mg\_grant} and \texttt{mg\_revoke} calls (see Table~\ref{syscalls}).

\subsection{\tool Threads}\label{labeling}
Each~\stool thread may have multiple~\tool attached to it. There is no concept of inheriting capabilities by default (e.g., in the style of \texttt{fork}) as this makes reasoning about security difficult~\cite{baumann2019fork}. 
Here, a tagged thread can create a child by calling \texttt{mg\_clone}; the child thread does not inherit any of its parent's capabilities. However, the parent can create a child with a list of its~\tool and selected capabilities as an argument of \texttt{mg\_clone}. 
For instance, in Figure~\ref{threading}, thread $3$ is a child of~\stool thread $2$, which only gets ``plus'' capabilities for both shared~\tool $18$ and $46$ via \texttt{mg\_clone} with a specific Label passed by its parent thread.

For a~\stool to propagate, it must be through transferring capabilities; this can be done directly by calling \texttt{mg\_transfer\_caps} for ``plus'' capabilities and \texttt{mg\_grant} for declassification. Both these operations are also possible via specific arguments of \texttt{mg\_clone} syscall when creating a child thread. Figure~\ref{threading} shows how each thread can use the~\tool API for creating tags, changing labels, and passing capabilities to other threads. 
For instance, thread $1$ gains access to~\stool $18$ by directly getting the $b^{+}$ capability from thread $2$. Since it does not have the $b^{-}$ capability, it cannot change~\stool $18$ permissions or its memory mappings.

Table~\ref{syscalls} describes the userspace~\stool API. A thread can create a tag by calling \texttt{mg\_alloc\_tag}, and the kernel will create and return a fresh unique tag. The thread that allocates a tag becomes its owner and can give the capabilities for the new tag to other threads.
Each thread specifies its security policies by mutating its labels via \texttt{mg\_modify\_label}, and can declassify its own~\tool via \texttt{mg\_declassify}.

Threads can lock access or permission changes of their~\tool via \texttt{mg\_lock}, which temporarily change~\stool tag to restrict any modifications of~\tool state. A locked~\stool can only be accessed by calling \texttt{mg\_unlock}.

\subsubsection{\tool Memory Management}

\input{utiles_api.tex}

To provide in-process isolation with good performance (\cref{dp}) we provide a virtual memory management abstraction within the kernel for~\tool-aware memory tagging, mappings, protection, page faults handling, and least privilege enforcement. This abstraction bypasses most of the kernel paging abstraction that improves its performance. Furthermore, it hides the intricacies of hardware domains (\cref{challenges}).
Then we provide a userspace library on top of our modified kernel, using our~\tool-specific system calls, for managing~\tool memory. 
An application creates a new~\stool by calling \texttt{mg\_create}; the kernel creates a unique tag with both capabilities (since it is the owner) and adds it to the thread's label and capability lists, and returns a unique ID. A~\stool can be kernel-backed (just depending on commodity pagetable for isolation) or hardware-backed which maps a~\stool to finer-grained memory safety/tagging features.
We extend the kernel VM layer to support~\tool and maintain a private per-\stool virtual page table (\texttt{pgd\_t}) that is loaded into the \texttt{TTBR} register when the thread needs to do memory operations inside an~\stool during a lightweight context switch. An internal~\stool data structure maintains its address space range and permissions as shown in the following code\cref{mgstruct}.

\begin{lstlisting}[caption={Internal~\stool data structure },captionpos=b,label={mgstruct}]
struct mg_struct {
    //operation bitmaps: set to 1 if mg[i] is allowed to do this operation, 0 OW
    DECLARE_BITMAP(mg_Read, MG_MAX); 
    DECLARE_BITMAP(mg_Write, MG_MAX);
    DECLARE_BITMAP(mg_Execute, MG_MAX); 
    DECLARE_BITMAP(mg_Allocate, MG_MAX);
    int mg_id;    
    struct mutex mg_mutex;
    struct mem_segment *mg_range;
   };
\end{lstlisting}

Threads (or Linux tasks) in a process share the same \texttt{mm\_struct} that describes the process address space. Having separate \texttt{mm\_struct} for threads would significantly impact system performance, as all the memory operations related to page tables should maintain strict consistency~\cite{hsu2016enforcing}. Instead, we extend \texttt{mm\_struct} to embed~\stool metadata within it as lightweight protected regions in the same address space as shown in \cref{mmstruct}.
It stores a per-\stool \texttt{pgd\_t} for threads and other metadata for memory management, fault handling, and synchronization.

The standard Linux kernel avoids reloading page tables during a context switch if two tasks belong to the same process. We modified \texttt{check\_and\_switch\_context}
to reload~\stool page tables and flush related TLB entries if one of the switching threads owns an~\stool.
We further mitigate the flushing overhead using ASID tagged TLB feature and ARM MDs.
We modify \texttt{mmap.c} to keep track of~\stool-mapped memory ranges and add \texttt{mg\_mmap/mumap} operations.

The kernel \texttt{handle\_mm\_fault} handler is also extended to specially manage page faults in~\stool regions, so an~\stool privilege violation results in the handler killing the violating thread.

\begin{lstlisting}[caption={Extending the Linux kernel mm\_struct with~\tool metadata.},captionpos=b,label={mmstruct}]
struct mm_struct {
...
#ifdef CONFIG_MG
    struct mg_struct *mg_metadata[MG_MAX];
    atomic_t num_mg;    /* number of mgs */
    pgd_t *mg_pgd_list[MG_MAX]; /*mg Page tables per threads.*/
    int curr_using_mg;
    spinlock_t sl_mg[MG_MAX];
    struct mutex mg_metadata_mut;
    DECLARE_BITMAP(mg_InUse, MG_MAX);
#endif
... };
\end{lstlisting}

Example code~\ref{api} shows a basic way of using~\tool to protect sensitive content in a single thread. 
Then the owner thread maps pages to its~\stool by calling \texttt{mg\_mmap} that updates the~\stool's metadata with its address space ranges. The kernel allows mappings based on the thread's labels and free hardware domains. If there is a free hardware domain, it maps pages to that domain and places it to~\tool cache. When the~\tool already exists in the cache, further access to it is fast. When there is no free hardware domain, we have to evict one of the~\tool from the cache and map the new~\stool metadata to the freed hardware domain; this requires storing all the necessary information for restoring the evicted~\stool, such as its permission, address space range, and tag. The caching process can be optimized by tuning the eviction rate and suitable caching policies similar to libMPK~\cite{park2018libmpk}.

The application uses \texttt{mg\_malloc} and the~\stool ID to allocate memory within the~\stool boundaries (mg\_malloc), and \texttt{mg\_free} to deallocate memory or \texttt{mg\_mprotect} to change its permissions  (see Table~\ref{umem}). The owner thread can use \texttt{mg\_lock} to restrict unauthorized access to it by accident or other malicious code; this is helpful for mitigating attacks inside a single thread. Then application developer can allow only his trusted functions or necessary parts of the code to gain access by calling \texttt{mg\_unlock} (e.g., our single-threaded OpenSSL use case in~\cref{openssl}). 
\input{code.tex}

Our current implementation of~\tool utilizes ARM-MDs for efficient in-process virtual memory tagging; as a result, only code running in supervisor mode can change a domain's access control via the DACR register (\S\ref{memdomain}) or remap private addresses to another domain through the TTBR domain bits. However, note that~\tool abstraction is designed to  support similar hardware memory tagging features such as MTE and PAC with straightforward changes; mostly by replacing the backed for~\tool memory management API (\texttt{mg\_malloc} layer) since the threading and other kernel changes are architecture-agnostic.
Our API and mappings prevent unauthorized permission changes for~\tool, and we also do not provide a userspace API for direct modification of the DACR.
Threads security policy enforcement is done by adding custom security hooks in the kernel's virtual memory management and task handling layers. It checks access based on the correct flow of threads labels (\S\ref{labeling}).
We extend the kernel page fault handler for~\tool-specific cases. Illegal access to~\tool causes domain faults which our handler logs (e.g., violating thread information) and terminates it with a signal.

\section{Implementation}\label{imp}

\textbf{\tool Kernel:}
The~\tool core access control enforcement and the security model is implemented in the form of a new Linux Security Module (LSM)~\cite{morris2002linux} with only four custom hooks. The LSM initializes the required data structures, such as the label registry and includes the implementation of all access control system calls (Table~\ref{syscalls}) for enforcing least privilege. This includes locking~\tool, changing labels, transferring capabilities,authority operations, and declassification based on the labeling mechanism (\S\ref{labeling}). 

We modify the Linux task structure to store the metadata required to distinguish~\tool tasks from regular ones.
Specifically, we add fields for storing~\tool metadata, label/ownership as an array data structure holding its tags (each tag is a 32-bit identification whose upper 2 bits stores plus and minus capabilities), a capability list; all included as task credential data structure. We implemented a hash table-based registry to make operations (e.g., store, set, get, remove) on these data structures more efficient.

 The LSM also provides custom security hooks for parsing userspace labels to the kernel (\texttt{copy\_user\_label}), labeling a task (\texttt{set\_task\_label}), checking whether the task is labeled (\texttt{is\_task\_labeled}), and checking if the information flow between two tasks is allowed (\texttt{check\_labels\_allowed}). These security hooks are added in various places within the kernel to~\tool are guarded against unauthorized access or permission change by either the POSIX API (e.g., mmap, mprotect, fork) or the~\tool API. For example, forking a labeled task should not copy its labels and capability lists, and this is enforced using the~\tool LSM hooks. As another example, to avoid a task performing unauthorized memory allocation into a random~\stool or mapping pages to it, the security hooks are in the kernel's virtual memory management layer where the~\tool memory management engine (Table~\ref{umem}) can enforce correct access.

The~\tool virtual memory abstraction is implemented as a set of kernel functions similar to their Linux equivalents (e.g., \texttt{do\_mmap}, \texttt{do\_munmap} and \texttt{do\_mprotect}) with similar semantics but with additional arguments that are required for enforcing the least privilege on~\tool.
 When an application creates a~\stool by calling \texttt{mg\_create} (or \texttt{mg\_mmap} for the first time), a~\stool ID passed as an argument that is associated with in-kernel metadata, together with the~\stool tag, and its capabilities that would be added to the task credentials. 
 
When~\tool are mapped to hardware domains, the exact physical domain number is hidden from the userspace code to avoid possible misuse of the API. The mappings between~\tool and hardware domains are maintained through a cache-like structure similar to libmpk~\cite{park2018libmpk}.  A~\stool is inside the cache if it is already associated with a hardware domain; otherwise, it evicts another~\tool based on the least recently used (LRU) caching policy while saving all require metadata for restoring the~\stool mapping and permission flags. 

Users can get their~\tool permissions by calling \texttt{mg\_get}, and quickly change its permission through \texttt{mg\_mprotect} if the requested permission change matches one of the domain's supported options (Table ~\ref{domains}) or undergo the small overhead of a dynamic security check otherwise.
Any violation of~\tool permissions causes a~\tool fault that leads to the violating thread being terminated. To protect~\tool against API attacks, all memory management system calls check whether the caller thread has the appropriate capabilities using the security hooks.

Creating a~\stool adds a new tag and owner capabilities to the task credential, and the userspace library also provides a management API for modifying labels and capabilities.  Each thread can use \texttt{mg\_transfer\_caps} for passing the plus capabilities to other threads, \texttt{mg\_grant\_revoke} for handling authorities, \texttt{mg\_lock} to prohibit access to a~\stool, and \texttt{mg\_unlock} to restore access. The \texttt{mg\_lock/unlock}  operations are helpful in limiting in-process buggy code from accessing~\tool content.

\textbf{Userspace:}
To reduce the size of the TCB, we did not modify existing system libraries and instead provided a userspace library to invoke~\tool system calls.
This library supports a familiar API for memory management within a~\stool, including \texttt{mg\_malloc} and \texttt{mg\_free} for memory management. We provide a custom memory allocator similar to HeapLayer~\cite{berger2001composing} that allocates memory from an already mapped~\stool. For each~\stool, there is a memory domain metadata structure that keeps essential information such as the~\stool address space range (base and length) and the two lists of free blocks from the head and tails of the~\stool region that is used when searching for free memory.

\section{Evaluation}\label{eval}

We evaluated our implementation of~\tool on a Raspberry Pi 3 Model B~\cite{rpi3} that uses a Broadcom BCM2837 SoC with a 1.2 GHz 64-bit quad-core ARM Cortex-A53 processor with $32KB$ L1 and $512KB$ L2 cache memory, running a 32-bit unmodified Linux kernel version 4.19.42 and glibc version 2.28 as the baseline.
We use microbenchmarks and modified applications to evaluate~\tool in terms of security, performance, and usability (\cref{dp} and \cref{memdomain}) by answering the following questions: 

\begin{itemize}[noitemsep]
    \item What is the initialization and runtime overhead of~\tool? How does using hardware domains impact performance?

    \item Are~\tool practical and adaptable for real-world applications? How much application change and programming effort is required? What is the performance impact? How does it perform in a multi-threaded environment?

   \item What is the memory footprint of~\tool? How much memory does it add (statically and dynamically) to both the kernel and userspace?

\end{itemize}

\subsection{Microbenchmarks}

\input{microbench.tex}

\textbf{Creating~\tool:}
Table~\ref{mmap} tests the cost of creating and mapping pages to~\tool using \texttt{mg\_mmap} when~\tool are directly mapped to hardware domains, 1MB aligned memory regions with only 16~\tool support, as compared to virtualized~\tool when there is no free hardware domain and requires evicting~\tool from the cache. The results show that the direct use of hardware domains improves~\tool performance by $4.9\%$ compare to the virtualized one. Note that creating~\tool is usually a one-time operation at the initial phase of an application.

  \begin{table}[H]
  \centering
 \resizebox{10cm}{!}{%

    \begin{tabular}{|l|l|l|}
    \hline
    Operation                                           & Overhead & stddev    \\ \hline
    Direct mg\_mmap/munmap   & 4.8\%                        & +- 0.17\% \\ \hline
    Virtualised mg\_mmap/munmap & 10.01\%                      & +- 0.15\% \\ \hline
    \end{tabular}}
    \caption{Cost of creating~\tool when directly mapped to hardware domains vs virualised mapping that requires~\tool caching. The results are average of 10000 runs.}
    \label{mmap}
\end{table}

\textbf{Memory protection \& allocation:}

We measure the cost of memory protection for baseline Linux where protection is per-process, and on \stool threads where protection is per-thread and either implemented in software or hardware.

Graph~\ref{mprot} shows the average results of $10000$ runs of our microbenchmark comparing the cost of \texttt{mg\_mprotect} with \texttt{mprotect} on baseline kernel. The results show \texttt{mg\_mprotect} is $1.12$x slower than \texttt{mprotect}, but the MD-backed \texttt{mg\_mprotect} is $1.14$x faster than baseline for some permissions (none and r/w) that supported by \texttt{DACR} register and do not need a TLB flush. Note that since hardware memory domains do not have flexible access control options, we cannot benefit from a control switch of domains using the DACR register for all possible permission flags such as the RO, WO, and EO variants.

\input{mprotect}

Allocating memory using \texttt{mg\_malloc} is on average 1.08x faster than glibc \texttt{malloc} for blocks $ \leq 64KB$ and introduces a small overhead ($8.3\%$) for blocks greater than $64KB$ (see Figure~\ref{memfig}). This cost can be optimised by using high-performance memory allocators. The results are average of running microbenchmarks 20000 times, and shows
using~\tool provides reasonable overhead for memory allocation and permission changes.

\textbf{Threading:}
We tested the cost of~\stool threading operations (creating and joining) through \texttt{mg\_clone} that creates~\stool-aware threads.  The test uses the \texttt{clone} syscall with minor modifications to restrict any credential sharing with the child by default (instead it provides additional clone options for passing parent's capabilities to its child). We implemented \texttt{mg\_join} using \texttt{waitpid}. Table~\ref{fork} shows \texttt{mg\_clone} outperforms \texttt{pthread\_create} by $0.56\%$ and \texttt{fork} by $83.01\% $.  This gain is attributed to the~\stool operations simply doing less work for initializing new threads.

\input{thrd.tex}

\textbf{Codebase overhead:}
Another factor towards the usability of~\tool is the size of the codebase, which is important both from a security perspective and the resource limitations of small devices. We implemented~\tool as a Linux kernel patch with no dependency on any userspace libraries. As Table~\ref{totalloc} shows it adds less than $5.5K$ LoC in total to both the kernel ($\approx 3K$ LoC) and userspace ($2.5K$ LoC).  It adds $7KB$ to the kernel image size and adds $204KB$ for kernel slabs at runtime. The userspace library only needs $\approx 10KB$ of memory. These results show the~\tool memory footprint is small and suitable for many resource-constrained uses.

   \begin{table}[t]
   \centering
\resizebox{12cm}{!}{%

    \begin{tabular}{|l|l|l|}
    \hline
    Overhead                & Linux Kernel               & Userspace     \\ \hline
    Added LoC             & 3023                       & 2405          \\ \hline
    Static Memory footprint & static(7KB) slab(204KB) & Static(10KB)  \\ \hline
    \end{tabular}}
    \caption{Memory overhead of~\tool in Linux Kernel and userspace}
    \label{totalloc}
\end{table}
 
 \subsection{OpenSSL}
 \label{openssl}

 Cryptographic libraries are responsible for securing all connected devices and network communication, yet have been a source or victim of severe vulnerabilities. Given these libraries' critical role, a single vulnerability can have a tremendous security impact. 
The well-known OpenSSL's Heartbleed vulnerability~\cite{durumeric2014matter}, for example, enabled attackers to access many servers' private data (up to $66\%$ of all websites were vulnerable). More recently, GnuTLS suffered a significant vulnerability allowing anyone to passively decrypt traffic (\cve{2020-13777}). Lazar et al.~\cite{lazar2014does} studied 269 cryptographic vulnerabilities, finding that only $17\%$ of the vulnerabilities they studied originated inside the cryptographic libraries, with the majority coming from improper uses of the libraries or interactions with other codebases. However, recent studies show that about $27\%$ of vulnerabilities in cryptographic software are cryptographic issues, and the rest are system-level issues, including memory corruption and interactions with the host or other applications/libraries~\cite{blessing2021you}. 

Hence, we modified OpenSSL to utilize~\tool for protecting private keys from potential information leakage by storing the keys in protected memory pages inside a single~\stool or multiple~\tool assigned per private key. Using multiple~\tool provides stronger security while adding more overhead due to the cost of caching~\tool.

To enable~\tool inside OpenSSL, all the data structures that store private keys such as \texttt{EVP\_PKEY} needed protected heap memory allocation. This meant replacing \texttt{OpenSSL\_malloc} wit \texttt{mg\_malloc} and using \texttt{mg\_mmap} at the initialization phase for creating one or multiple (per session)~\tool to store private keys. After storing the keys, access to~\tool is disabled by calling \texttt{mg\_lock}. Only trusted functions that require access to private keys (e.g., \texttt{EVP\_EncryptUpdate} or \texttt{pkey\_rsa\_encrypt/decrypt}) can access~\tool by calling \texttt{mg\_unlock}. Modifying OpenSSL required fairly small code changes, and added 281 lines-of-code.

%    \begin{tabular}{|l|l|l|}
%    \hline
%    Overhead                & OpenSSL               & LevelDB     \\ %\hline
%    Added LoC               & 281                       & 157          \\ \hline
%    \end{tabular}
%    \caption{Programming effort of enabling~\tool in OpenSSL and %LevelDB}

We measured the performance overhead of~\tool-enabled OpenSSL by evaluating it on the Apache HTTP server (httpd) that uses OpenSSL to implement HTTPS. Table~\ref{httpd} shows the overhead of ApacheBench httpd with both the original OpenSSL library and the secured one with~\tool. ApacheBench is launched 100 times with various request parameters. We choose the TLS1.2 DHE-RSA-AES256-GCM-SHA384 algorithm with 2048-bit keys as a cipher suite in the evaluation.

The results show that on average~\tool introduces $0.47\%$  performance overhead in terms of latency when using a single~\stool for protecting all keys, and $3.67\%$ overhead when using a separate~\stool per session key. In the single~\stool case, the negligible overhead is mainly caused by in-kernel data structure maintenance for enforcing privilage separation and handling~\tool metadata. In the multiple-\tool case, since httpd utilizes more than 16~\tool (allocates a new~\stool per session), it causes higher overhead due to the caching costs within the kernel. 
 
\input{openssl.tex}
 
\subsection{LevelDB}\label{leveldb} 

Google's LevelDB is a fast key-value store and storage engine used by many applications as a backend database. It supports multithreading for both concurrent writers to safely insert data into the database as well as concurrent read to improve its performance. However, there is no privilege separation between threads, so each could have its private content isolated from other threads. We modified LevelDB to evaluate performance overhead of using the~\tool threading model when each thread has its own private storage that cannot be accessed by other threads.

We replaced the LevelDB threading backend (\texttt{env\_posix}) that uses pthreads with~\tool-aware threading, where each thread creates an isolated~\stool as its private storage and computation. We used the LevelDB \texttt{db\_bench} tool (without modification) for measuring the performance overhead of~\tool.

We generate a database with 400K records with 16-byte keys and 100-byte values (a raw size of 44.3MB). The number of reader threads is set to 1, 2, 4, 8, 16,  and 32 threads for each successive run. The threads operate on randomly selected records in the database. The results in Figures~\ref{ldb1} and ~\ref{ldb2} show how multithreading can improve the performance of LevelDB, and utilising~\tool adds a small overhead on write ($5\%$) and read ($1.98\%$) throughput.  As with OpenSSL previously, modifying LevelDB required only adding 157 lines-of-code around the codebase.

 \input{leveldb.tex}

 \section{Discussion \& Conclusion}\label{diss}
We have shown that~\tool provides a practical and efficient mechanism for intra-process isolation and inter-thread privilege separation on data objects. It adds small performance overhead and minimal memory footprint, which in essential for mobile and resource-constrained devices. However, the mechanism can still be taken further.

\subsection{Address Space Protection Limitations}

For single-threaded scenarios (e.g., event-driven servers), although~\tool can protect sensitive content from unsafe libraries or untrusted parts of the applications, it can be vulnerable if the untrusted modules are also ~\tool-aware and already use the~\tool APIs.  The application can use \texttt{mg\_get} to query~\stool information and use the API to access them. This is not an issue when the untrusted code is running in a separate thread since the kernel does not provide it the capabilities required for accessing the other~\tool.  It should be possible to modify popular event-driven libraries (e.g., libuv) to use threads purely to separate sensitive information such as key material, but we have not yet implemented this.

Various covert attacks~\cite{sigurbjarnarson2018nickel} and side-channel attacks such as Meltdown~\cite{lipp2018meltdown} and Spectre~\cite{kocher2018spectre} demonstrate how hardware and kernel isolation can be bypassed~\cite{hunt2019isolation}.~\tool are currently vulnerable to these class of attacks, although the existing countermeasures within the Linux kernel are sufficient protection. We believe these types of attacks are important security threats, and hardening~\tool against them could be significant future work.

\subsection{Compatibility Limitations}

Providing a solution that is compatible with various operating systems and heterogeneous hardware is challenging. Though we picked our base kernel on Linux and built the abstraction with minimal dependencies, some application modification is still required.  We believe that building more compatibility layers into our existing userspace implementation is possible and are open-sourcing our code to gather further feedback and patches from the relevant upstream projects we have modified.

Although Linux is the most widespread general-purpose kernel for embedded devices, as well as being the base for Android, still many even smaller devices depend on operating systems such as FreeRTOS. These often use ARM Cortex-M based hardware features for isolation (such as memory protection units (MPUs)~\cite{mpu,cmsis}), or more modern CPUs with memory tagging extension~\cite{mte}. We plan to explore the implementation of the~\tool kernel memory management on these single-address space operating systems, as well as broadening the port to Intel and PowerPC architectures on Linux (where the memory domains support is generally simpler to use than on ARM).

\section{Related work}\label{related}
There are many software or hardware-based techniques for providing process and in-process memory protection. 

\textbf{OS/hypervisor-based solutions:}
Hardware virtualization features are used for in-process data encapsulation by Dune~\cite{belay2012dune} by using the Intel VT-x virtualization extensions to isolate compartments within user processes. However, overall, the overheads of such virtualization-based encapsulation are more heavy-weight than~\tool. 
ERIM~\cite{vahldiek2019erim}, light-weight contexts (lwCs)~\cite{litton2016light} and secure memory views (SMVs)~\cite{hsu2016enforcing} all provide in-process memory isolation and have reduced the overhead of sensitive data encapsulation on x86 platforms.
The~\tool provides stronger security guarantees and privilege separation, allows more flexible ways of defining security policies for legacy code -- e.g., without the use of threads as in our OpenSSL example, its small memory footprint makes it suitable for smaller devices, and it takes advantage of efficient virtual memory tagging by using hardware domains to reduce overhead.
Burow et al.~\cite{burow2019sok} leverage the Intel MPK and memory protection extensions (MPX) to efficiently isolate the shadow stack. Our efforts to provide an OS abstraction for in-process memory protection is orthogonal to these studies, which all have potential use cases for~\tool.  Our focus has also been on lowering the resource cost to work well on embedded and IoT devices, while these projects are also currently x86-only.
HiStar~\cite{zeldovich2006making} is a DIFC-based OS that supports fine-grained in-process address space isolation, which influenced our work, but we focused on providing a more general-purpose solution for small devices by basing our work on the Linux kernel instead of a custom operating system. Flume~\cite{krohn2007information} proposed process-level DIFC as a minimal extension to the Linux kernel, making DIFC work with the languages, tools, and OS abstractions already familiar to programmers. It also introduced a cleaner label system (which HiStar have later adopted). 
Likewise, other DIFC-based systems only support per-process protection. They also add large overhead~\cite{wang2015between,krohn2007information} or need specific programming language support~\cite{roy2009laminar}. \tool, however, do no aim to enforce dataflow protection on all system objects, but only focuses on threads and address space objects to enable very lightweight privilege separation.

\textbf{Compiler \& Language Runtime:}
Various compiler techniques introduce memory isolation as part of a memory-safe programming language. These approaches are fine-grained and efficient if the checks can be done statically~\cite{elliott2018checked}. However, such isolation is language-specific, relies on the compiler and runtime, and not effective when applications are co-linked with libraries written in unsafe languages.~\tool abstractions are fine-grained enough to be useful to these tools, for example, to isolate unsafe bindings.
Software fault isolation (SFI)~\cite{wahbe1994efficient,sehr2010adapting} uses runtime memory access checks inserted by the compiler or by rewriting binaries to provide memory isolation in unsafe languages with substantial overhead. Bounds checks impose overhead on the execution of all components (even untrusted ones), and additional overhead is required to prevent control-flow hijacks, which could bypass the bounds checks~\cite{koning2017no}.
ARMLock~\cite{zhou2014armlock} is an SFI-based solution that offers lower overhead utilizing ARM MDs.
Similarly, Shreds~\cite{chen2016shreds} provides new programming primitives for in-process private memory support.
\tool also uses ARM MDs for improving the performance of intra-process memory protection, but is a more flexible solution for intra-process privilege separation; it provides a new threading model for dynamic fine-grained access control over the address space with no dependency on a binary rewriter, specific compiler or programming language.

\textbf{Hardware-enforced techniques:}
A wide range of systems use hardware enclaves/TEEs such as Intel's SGX~\cite{anati2013sgx} or ARM's TrustZone~\cite{tzcm} to provide a trusted execution environment for applications that against malicious kernel or hypervisor~\cite{guan2017trustshadow,frassetto2017jitguard,arnautov2016scone,tarkhani2019snape,mo2022sok}.
The trust model exposed by these hardware features is very fixed, and usually results in porting monolithic codebases to execute within the enclaves. Hence, there are wide ranges of attack vectors, which many are memory vulnerabilities inside enclaves or their untrusted interface, in such systems~\cite{singh2021enclaves,tarkhani2022secure}. EnclaveDom~\cite{melara2019enclavedom} utilizes Intel MPK to provide in-enclave privilege separation.~\tool provide better performance and more general solutions with no dependency on these hardware features; hence it can be used for in-enclave isolation and secure multi-threading to improves both security and performance of enclave-assisted applications~\cite{tarkhani2020enclave}.
Ultimately, dedicated hardware support for tagged memory and capabilities would be the ideal platform to run~\tool on~\cite{zeldovich2008hardware}. We are planning on supporting more of these hardware features as future work, with a view to analyzing if the overall increase in hardware complexity offsets the resource usage in software for embedded systems.

\bibliographystyle{splncs04}
\bibliography{ref}

\end{document}

%% file: cve_tab.tex
\begin{table}[htb]
\centering
\resizebox{12cm}{!}{%
  \footnotesize
  \begin{tabular}{|c|l|l|c|}
      \hline

    & \bf example CVE       & \bf Description  & \bf \tool \\
    \hline
    \parbox[t]{2mm}{\multirow{7}{*}{\rotatebox[origin=c]{90}{\bf In-Process threats}}}
      &\cve{2021-3450}    & Improper access control in shared library & \chk \\

  &\cve{2021-29922}    & unsafe language binding & \chk \\
        &  \cve{2021-31162}  & Rust runtime memory  corruption & \chk \\  
    & \cve{2019-9345} & Shared mapping bug  &\chk \\
    & \cve{2021-45046} & thread-based privilege escalation & \chk \\
    & \cve{2019-9423} & missing bounds check      & \chk \\
    & \cve{2019-15295} & unsafe third party library      & \chk \\
    & \cve{2019-1278} &  unsafe third party library   &  \chk\\
    & \cve{2018-0487} & unsafe third party library    & \chk \\
    
    & \cve{2017-1000376} &  unsafe native bindings     & \chk \\

    & \cve{2014-0160} &  Heartbleed bug     & \chk \\ 
        & \cve{2021-3177} & Python ctypes memory leak & \chk \\
        & \cve{2021-28363} & Python ctypes memory leak & \chk \\

    \hline
    
        \parbox[t]{2mm}{\multirow{3}{*}{\rotatebox[origin=c]{90}{\bf Other}}}

                &  && \\

     & \cve{2018-0497} & SW side-channels  &   \\
      & \cve{2017-5754} & HW side-channels & \\

    \hline

  \end{tabular}}
  \caption{\label{t:cve-table}A representative selection of vulnerabilities that cause sensitive content leakage. The attacks with a tick can be mitigated by using {\bf \tool} protection.}
  \label{cvetab}
\end{table}

%% file: domains.tex
\begin{table}[htb]
\centering
\resizebox{10cm}{!}{%

    \begin{tabular}{|l|l|l|}
    
    \hline
    Mode & Bits & Description                                                                     \\ \hline
    No Access   & 00 &  Any access causes a domain fault.                          \\\hline
    Manager     & 11 & Full accesses with no permissions check.\\\hline
    Client      & 01 &  Accesses are checked against the page tables   \\\hline
    Reserved    & 10 &   Unknown behaviour. \\ \hline
    \end{tabular}}
    \caption{ARM memory domains access permissions }
    \label{domains}
\end{table}

%% file: utiles_acc.tex
\begin{table}[t]
\centering
\resizebox{13cm}{!}{%

    \begin{tabular}{|l|l|}
    \hline
    syscalls                        &  Description                                                                                                \\ \hline
    mg\_alloc\_tag()$\to$ $t$                  & allocate a unique tag \\ \hline
    mg\_modify\_label($L$)                    & modify a thread's label/tag    \\ \hline
    mg\_transfer\_caps($L\to{c*}, p$)         & passing capabilities to thread $p$\\ \hline
    mg\_declassify($L\to{t*}$)           & thread declassification or endorsement   \\ \hline
    mg\_grant($L\to{t*}, p1, p2$)          & adds an acts-for or a delegation link \\ \hline
    mg\_revoke\_grant($L\to{t*}, p1, p2$)    & removes an acts-for or a delegation link \\ \hline
    mg\_lock ($L\to{t*}$)         & disables access to an object  \\ \hline
    mg\_unlock ($L\to{t*}$) & enables access to a locked object \\ \hline
    mg\_clone ($L,int(*fn)(void*)...$) $\to$ $p$      & creates a thread     \\ \hline

    \end{tabular}}
  \caption{\tool access control system calls. $Pi$ represents principal $i$, $L$ as a label that is a list of tags ($t*$) and their capabilities ($c*$). }  
\label{syscalls} 

\end{table}

%% file: utiles_api.tex
   \begin{table}[t]
      \centering

\resizebox{12cm}{!}{%

    \begin{tabular}{|l|l|}
    \hline
     Name    & Description                                 \\ \hline
     
    mg\_create $\to$ {id}    & Create a new~\stool                       \\ \hline
    
    mg\_kill($id$)    & Destroy a~\stool                       \\ \hline
    
    mg\_malloc($id,size$) $\to$ {void*}    & Allocate memory within a~\stool  \\ \hline
    
    mg\_free($id,void*$)     & free memory from a ~\stool                        \\ \hline
    
    mg\_mprotect($id,...$) & change an~\stool's pages permission      \\ \hline
    
    mg\_mmap($id,...$)$\to$ {void*} & Map a page group to a~\stool      \\ \hline
    
    mg\_munmap($id,...$) & Unmap all pages of a~\stool     \\ \hline
    
    mg\_get($id$)$\to$ {$perms$}     & Get a~\stool permission      \\ \hline
    \end{tabular}
    }
    \caption{Some of userspace \tool memory management API. Each~\stool has an $id$ and is a tagged kernel object internally. \tool access control is checked within the kernel.}
    \label{umem}
\end{table}

%% file: code.tex
\begin{lstlisting}[caption={Basic \tool usage},captionpos=b,label={api}]
    /* create a microgaurd (i.e., mg_id) */
    int mg_id = mg_create(); 
    
    /* map a memory region to the mg */
    memblock = (char*) mg_mmap(mg_id, addr, len, prot , 0, 0); //
    
    // set permissions by mg_mprotect

    /* allocate memory from mg */
    private_blk = (char*) mg_malloc(mg_id, priv_len);

    /* make mg inaccessible */
    lock_mg(mg_id);

    //... untrusted computations ....//

    /* make mg accessible */
    unlock_mg(mg_id);

    //... trusted computations ....//

    /* cleanup mg */
    mg_free(private_blk);
    mg_munmap(mg_id, memblock,len);

\end{lstlisting}

%% file: microbench.tex
 \begin{figure}[t]
 \centering
\resizebox{12cm}{!}{%
 \begin{tikzpicture}
    \begin{axis}[
        width=0.6\textwidth,
        height=4.5cm,
        bar width=7pt,
        % commented to see, if the result is correct
%        major x tick style=transparent,
        symbolic x coords={1,2,4,8,16,32,64,128,256,512},
        xlabel={Allocated Memory (KB)},
        xtick=data,
        %x label style={at={(axis description cs:0.5,-0.1)},anchor=north},
        % changed from `\pgflinewidth' to 1 easier see that the result is correct
        ymin=0,
         ymax=15,
        % changed to 0 to make the computation of the shift easier
        ybar=0pt,
        ymajorgrids=true,
        ylabel={Latency(ms)},
        legend cell align=left,
        legend style={
            at={(1,1.05)},
            anchor=south east,
            column sep=1ex
        },
    ]
        % add a scope around the to shift bars
        \begin{scope}[
            % shift the bars accordingly
            xshift={0.1*\pgfplotbarwidth},
            % and just "fill" the bars
            % (to avoid overlapping of the more right bars to the more left ones)
            draw=none,
        ]
            % your style could heavily be simplified to just providing the `+'
            % sign and the color as option
            %\addplot+ [bblue]  coordinates
             %   {(St.~2,10006) (St.~3,99895) (St.~4, 99867)};
            %\addplot+ [rred]   coordinates
             %   {(St.~1,1) (St.~2,10006) (St.~3,99448) (St.~4, 99487)};

\addplot+ [draw=none, fill=red!60,
            error bars/.cd,
                y dir=both,
                % (changed from `y explicit` so the error bars are (clearly) visible
                y explicit relative,
        ] coordinates {
            (1,10.12) +- (0,0.09)
            (2,9.80) +- (0,0.1)
            (4,10.12) +- (0,0.09)
            (8,10.08) +- (0,0.1)
            (16,9.66) +- (0,0.1)
            (32,9.42) +- (0,0.1)
            (64,10.61) +- (0,0.08)
            (128,10.18)  +- (0,0.09)
            (256,10.47) +- (0,0.09)
            (512,10.43) +- (0,0.09)
        };
        
        \addplot+ [draw=none, blue!60,
            error bars/.cd,
                y dir=both,
                % (changed from `y explicit` so the error bars are (clearly) visible
                y explicit relative,
        ] coordinates {          
            (1,9.13) +- (0,0.13)
            (2,9.12) +- (0,0.13)
            (4,8.7) +- (0,0.13)
            (8,9.04) +- (0,0.13)
            (16,9.15) +- (0,0.12)
            (32,9.03) +- (0,0.13)
            (64,10.44) +- (0,0.09)
            (128,10.27)  +- (0,0.1)
            (256,11.30) +- (0,0.08)
            (512,12.12) +- (0,0.09)
                        };

        \end{scope}

\legend{malloc \& free,mg\_malloc \& mg\_free}
    \end{axis}
\end{tikzpicture}}
\caption{Cost of \tool memory allocation (malloc \& free). On average \texttt{mg\_malloc} outperforms \texttt{malloc} by a small rate ($0.03\%$).}

\label{memfig}
 \end{figure}
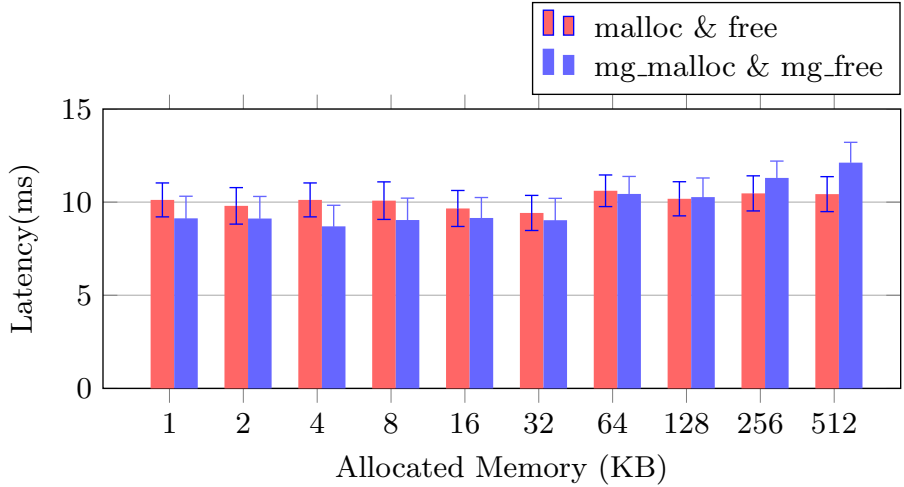

%% file: mprotect.tex
\begin{figure}[htp]
\centering
\resizebox{12cm}{!}{%
 \begin{tikzpicture}
    \begin{axis}[
        width=0.6\textwidth,
        height=4cm,
        bar width=7pt,
        % commented to see, if the result is correct
%        major x tick style=transparent,
        symbolic x coords={RO, WO, EO, R/W, NO-ACC},
        xtick=data,
        %x label style={at={(axis description cs:0.5,-0.1)},anchor=north},
        % changed from `\pgflinewidth' to 1 easier see that the result is correct
        ymin=0,
         ymax=3.5,
        % changed to 0 to make the computation of the shift easier
        ybar=0pt,
        ymajorgrids=true,
        ylabel={Latency(\textmu s)},
        legend cell align=left,
        legend style={
            at={(1,1.05)},
            anchor=south east,
            legend columns=3,
            column sep=1ex
        },
    ]
        % add a scope around the to shift bars
        \begin{scope}[
            % shift the bars accordingly
            xshift={0.1*\pgfplotbarwidth},
            % and just "fill" the bars
            % (to avoid overlapping of the more right bars to the more left ones)
            draw=none,
        ]
            % your style could heavily be simplified to just providing the `+'
            % sign and the color as option
            %\addplot+ [bblue]  coordinates
             %   {(St.~2,10006) (St.~3,99895) (St.~4, 99867)};
            %\addplot+ [rred]   coordinates
             %   {(St.~1,1) (St.~2,10006) (St.~3,99448) (St.~4, 99487)};

        \addplot+ [draw=none, blue!30,
            error bars/.cd,
                y dir=both,
                % (changed from `y explicit` so the error bars are (clearly) visible
                y explicit relative,
        ] coordinates {          
            (RO,2.239) +- (0,0.01)
             (WO,2.343) +- (0,0.1)
             (EO,2.244) +- (0,0.02)
            (R/W,2.89) +- (0,0.03)
            (NO-ACC,2.234)  +- (0,0.1)
                        };

        \addplot+ [draw=none, blue!60,
            error bars/.cd,
                y dir=both,
                % (changed from `y explicit` so the error bars are (clearly) visible
                y explicit relative,
        ] coordinates {          
            (RO,2.91) +- (0,0.13)
             (WO,2.53) +- (0,0.11)
             (EO,2.58) +- (0,0.12)
            (R/W,2.96) +- (0,0.13)
            (NO-ACC,2.53)  +- (0,0.12)
                        };

                \addplot+ [draw=none, green!40,
            error bars/.cd,
                y dir=both,
                % (changed from `y explicit` so the error bars are (clearly) visible
                y explicit relative,
        ] coordinates {          
            (RO,2.932)  +- (0,0.13)
            (WO,2.609) +- (0,0.12)
            (EO,2.598) +- (0,0.11)
            (R/W,2.437) +- (0,0.13)
            (NO-ACC,2.026) +- (0,0.12)

                        };
        \end{scope}

\legend{mprotect,mg\_mprotect,hw mg\_mprotect}
    \end{axis}
\end{tikzpicture}}
%\caption{Average cost of mg permission changes.}
\label{mprot}
\end{figure}

%% file: thrd.tex
  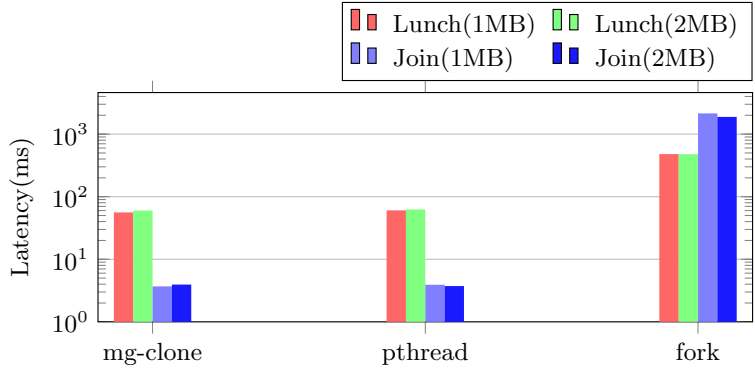
\begin{figure}[t]
  \centering
\resizebox{10cm}{!}{%

 \begin{tikzpicture}
    \begin{axis}[
        width=0.6\textwidth,
        height=4.5cm,
        bar width=7pt,
        % commented to see, if the result is correct
%        major x tick style=transparent,
        symbolic x coords={mg-clone,pthread,fork},
        xlabel={},
        xtick=data,
        %x label style={at={(axis description cs:0.5,-0.1)},anchor=north},
        % changed from `\pgflinewidth' to 1 easier see that the result is correct
        ymin=1,
        % changed to 0 to make the computation of the shift easier
        ybar=0pt,
        ymajorgrids=true,
        ylabel={Latency(ms)},
        ymode=log,
         legend cell align=left,
        legend style={
            at={(1,1.05)},
            anchor=south east,
            legend columns=2,
            column sep=1ex
        },
    ]
        % add a scope around the to shift bars
        \begin{scope}[
            % shift the bars accordingly
            % and just "fill" the bars
            % (to avoid overlapping of the more right bars to the more left ones)
            draw=none,
        ]

        \addplot [draw=none, fill=red!60] coordinates {
        (mg-clone,56.099) 
        (pthread,60.439)
        (fork,479.231) };
         \addplot [draw=none, fill=green!50] coordinates {
          (mg-clone,60.167) 
          (pthread,62.460)
          (fork,477.621) };
          
         \addplot [draw=none, fill=blue!50] coordinates {
         (mg-clone,3.688) 
         (pthread,3.914)
         (fork,2143) };

           \addplot [draw=none, fill=blue!90] coordinates {
           (mg-clone,3.941) 
           (pthread,3.731)
           (fork,1884) };

        \end{scope}

\legend{Lunch(1MB),Lunch(2MB),Join(1MB), Join(2MB) }
    \end{axis}
\end{tikzpicture}}
\caption{Overhead of creating \stool-enabled threads: the results are the average of 100000 runs with 1MB and 2MB heap sizes. On average, \texttt{mg\_clone} latency is $5.39\%$ lower than of \texttt{pthread\_create}. }
\label{fork}
 \end{figure}

%% file: openssl.tex
   \begin{figure}[htb]
   \centering
\resizebox{14cm}{!}{%

 \begin{tikzpicture}
    \begin{axis}[
        width=0.7\textwidth,
        height=4.5cm,
        bar width=7pt,
        % commented to see, if the result is correct
%        major x tick style=transparent,
        symbolic x coords={1,2,4,8,16,32,64,128,256, 512},
        xlabel={number of requests},
        xtick=data,
        %x label style={at={(axis description cs:0.5,-0.1)},anchor=north},
        % changed from `\pgflinewidth' to 1 easier see that the result is correct
        ymin=1,
        % changed to 0 to make the computation of the shift easier
        ybar=0pt,
        ymajorgrids=true,
        ylabel={Latency(s)},
        ymode=log,
        legend cell align=left,
        legend style={
            at={(1,1.05)},
            anchor=south east,
            column sep=1ex
        },
    ]
        % add a scope around the to shift bars
        \begin{scope}[
            % shift the bars accordingly
            % and just "fill" the bars
            % (to avoid overlapping of the more right bars to the more left ones)
            draw=none,
        ]

\addplot  [draw=none, fill=blue!40] coordinates { (1,0.618) (2,1.396) (4,2.932) (8,5.260) (16,10.794) (32,22.569) (64,44.525) (128,95.875) (256,180.768) (512,365.445)  };
\addplot  [draw=none, fill=green!60] coordinates { (1,0.750) (2,1.442) (4,2.694) (8,5.371) (16,10.627) (32,21.275) (64,45.495) (128,94.714)(256, 188.252) (512,370.352)  };
\addplot  [draw=none, fill=red!60] coordinates { (1,0.782) (2,1.278) (4,2.776) (8,5.246) (16,10.451) (32,22.861) (64,49.201 ) (128,98.023) (256, 209.613)(512,386.236)  };

        \end{scope}

\legend{original,\tool (single \stool), \tool (per session \tool)}
    \end{axis}
\end{tikzpicture}}
\caption{Overhead of httpd on unmodified OpenSSL vs \tool-enabled one.}
%overhead of using single \stool for protecting private keys is $0.47\%$ while using multiple \tool (one per session) causes a larger overhead of $3.67\%$ compare to using unmodified OpenSSL baseline.}
\label{httpd}
 \end{figure}
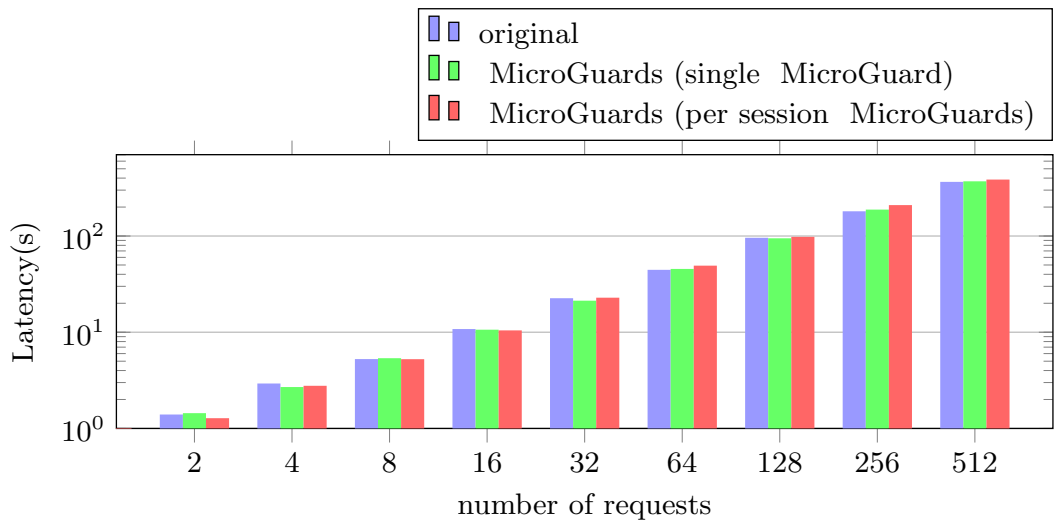

%% file: leveldb.tex
 
   \begin{figure}[htb]
   \centering
\resizebox{10cm}{!}{%

 \begin{tikzpicture}
    \begin{axis}[
        width=0.5\textwidth,
        height=4cm,
        bar width=8pt,
        % commented to see, if the result is correct
%        major x tick style=transparent,
        symbolic x coords={1,2,4,8,16,32},
        xlabel={Number of Threads},
        xtick=data,
        %x label style={at={(axis description cs:0.5,-0.1)},anchor=north},
        % changed from `\pgflinewidth' to 1 easier see that the result is correct
        ymin=1,
        % changed to 0 to make the computation of the shift easier
        ybar=0pt,
        ymajorgrids=true,
        ylabel={Write Throughput(MB/s)},
        legend cell align=left,
        legend style={
            at={(1,1.05)},
            anchor=south east,
            column sep=1ex
        },
    ]
        % add a scope around the to shift bars
        \begin{scope}[
            % shift the bars accordingly
            % and just "fill" the bars
            % (to avoid overlapping of the more right bars to the more left ones)
            draw=none,
        ]

\addplot  [draw=none, fill=blue!90] coordinates { (1,1.3) (2,1.3) (4,1.5) (8,1.4) (16,1.2) (32,1.3 )   };
\addplot  [draw=none, fill=red!65 ] coordinates { (1,1.2) (2,1.4) (4,1.3) (8,1.4) (16,1.1) (32, 1.2) };

        \end{scope}

\legend{original, \tool}
    \end{axis}
\end{tikzpicture}}
\caption{LevelDB: performance overhead of\tool-based multithreading compare to pthread-based in terms of write throughput ($5\%$).}
\label{ldb1}
 \end{figure}
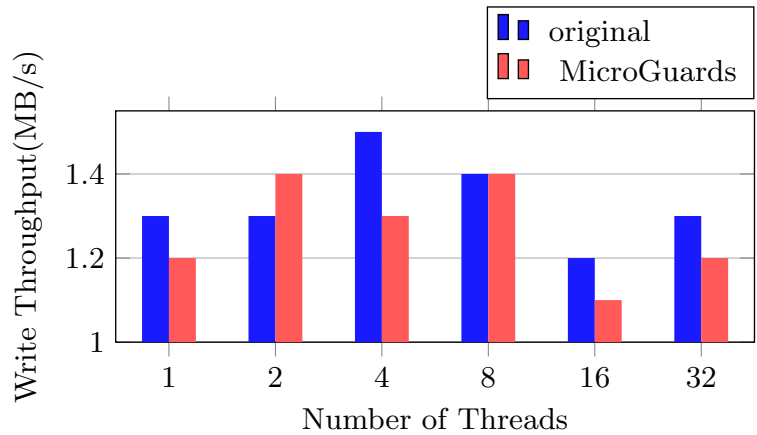

    \begin{figure}[htb]
    \centering
\resizebox{10cm}{!}{%

 \begin{tikzpicture}
    \begin{axis}[
        width=0.5\textwidth,
        height=4cm,
        bar width=8pt,
        % commented to see, if the result is correct
%        major x tick style=transparent,
        symbolic x coords={1,2,4,8,16,32},
        xlabel={Number of Threads},
        xtick=data,
        %x label style={at={(axis description cs:0.5,-0.1)},anchor=north},
        % changed from `\pgflinewidth' to 1 easier see that the result is correct
        ymin=1,
        % changed to 0 to make the computation of the shift easier
        ybar=0pt,
        ymajorgrids=true,
        ylabel={Read Throughput(MB/s)},
        legend cell align=left,
        legend style={
            at={(1,1.05)},
            anchor=south east,
            column sep=1ex
        },
    ]
        % add a scope around the to shift bars
        \begin{scope}[
            % shift the bars accordingly
            % and just "fill" the bars
            % (to avoid overlapping of the more right bars to the more left ones)
            draw=none,
        ]

\addplot  [draw=none, fill=blue!45] coordinates { (1,78.4) (2,80.2) (4,150.3) (8,268.5) (16,300.5) (32,280.6)  };

\addplot  [draw=none, fill=green!90] coordinates { (1,80.4) (2,79.0) (4,153.3) (8,299.3) (16,301.6) (32, 267.9)};

        \end{scope}

\legend{original, \tool}
    \end{axis}
\end{tikzpicture}}
\caption{LevelDB: performance overhead of\tool-based multithreading compare to pthread-based in terms of read throughput ($1.98\%$).}
\label{ldb2}
 \end{figure}
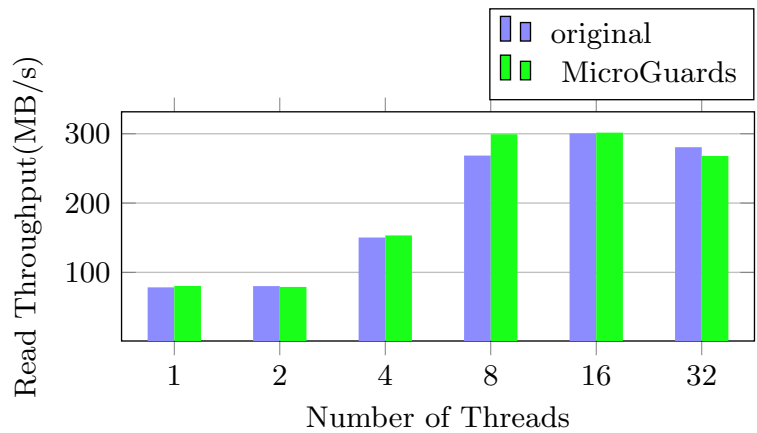
 
 

%% file: main.bbl
\begin{thebibliography}{10}
\providecommand{\url}[1]{\texttt{#1}}
\providecommand{\urlprefix}{URL }
\providecommand{\doi}[1]{https://doi.org/#1}

\bibitem{tserver}
Format string vulnerability in the {Cherokee}.
  \url{https://www.cvedetails.com/cve/CVE-2004-1097/}, access Date : 2020-1-5

\bibitem{iot19}
Iot developer survey 2019.
  \url{https://iot.eclipse.org/resources/iot-developer-survey/iot-developer-survey-2019.pdf}

\bibitem{rpi3}
Raspberry {Pi} 3 model b.
  \url{https://www.raspberrypi.org/products/raspberry-pi-3-model-b}

\bibitem{breach3}
Cyber security breaches survey 2018.
  \url{https://www.gov.uk/government/statistics/cyber-security-breaches-survey-2018}
  (2018)

\bibitem{breach1}
List of data breaches.
  \url{https://en.wikipedia.org/wiki/List_of_data_breaches} (2018)

\bibitem{almohri2018fidelius}
Almohri, H.M., Evans, D.: Fidelius charm: Isolating unsafe rust code. In:
  Proceedings of the Eighth ACM Conference on Data and Application Security and
  Privacy. pp. 248--255. ACM (2018)

\bibitem{anati2013sgx}
Anati, I., Gueron, S., Johnson, S., Scarlata, V.: Innovative technology for
  {CPU} based attestation and sealing. In: Proceedings of the 2nd international
  workshop on hardware and architectural support for security and privacy.
  vol.~13. ACM New York, NY, USA (2013)

\bibitem{cmsis}
ARM: Cmsis-zone.
  \url{https://arm-software.github.io/CMSIS_5/Zone/html/index.html}

\bibitem{arm2012architecture}
ARM: Architecture reference manual; {ARM}v7-a and {ARM}v7-r edition.
  \url{https://static.docs.arm.com/ddi0406/c/DDI0406C_C_arm_architecture_reference_manual.pdf}
  (2012), access Date : 2020-5-26

\bibitem{tzcm}
ARM: {ARM®v8-M Security Extensions:} requirements on development tools (2015)

\bibitem{mte}
ARM: {ARM} architecture reference manual {Armv8}, for {Armv8-A} architecture
  profile documentation. \url{https://developer.arm.com/docs/ddi0487/latest}
  (2018), access Date : 2020-5-26

\bibitem{arnautov2016scone}
Arnautov, S., Trach, B., Gregor, F., Knauth, T., Martin, A., Priebe, C., Lind,
  J., Muthukumaran, D., O'keeffe, D., Stillwell, M., et~al.: {SCONE}: Secure
  {Linux} containers with {Intel SGX}. In: OSDI. vol.~16, pp. 689--703 (2016)

\bibitem{baumann2019fork}
Baumann, A., Appavoo, J., Krieger, O., Roscoe, T.: A fork () in the road. In:
  Proceedings of the Workshop on Hot Topics in Operating Systems. pp. 14--22.
  ACM (2019)

\bibitem{belay2012dune}
Belay, A., Bittau, A., Mashtizadeh, A., Terei, D., Mazi{\`e}res, D., Kozyrakis,
  C.: Dune: Safe user-level access to privileged cpu features. In: Presented as
  part of the 10th {USENIX} Symposium on Operating Systems Design and
  Implementation ({OSDI} 12). pp. 335--348 (2012)

\bibitem{berger2001composing}
Berger, E.D., Zorn, B.G., McKinley, K.S.: Composing high-performance memory
  allocators (2001)

\bibitem{bittau2008wedge}
Bittau, A., Marchenko, P., Handley, M., Karp, B.: Wedge: Splitting applications
  into reduced-privilege compartments. In: USENIX Association (2008)

\bibitem{blessing2021you}
Blessing, J., Specter, M.A., Weitzner, D.J.: You really shouldn't roll your own
  crypto: An empirical study of vulnerabilities in cryptographic libraries.
  arXiv preprint arXiv:2107.04940  (2021)

\bibitem{brumley2004privtrans}
Brumley, D., Song, D.: Privtrans: Automatically partitioning programs for
  privilege separation. In: USENIX Security Symposium. pp. 57--72 (2004)

\bibitem{burow2019sok}
Burow, N., Zhang, X., Payer, M.: Sok: Shining light on shadow stacks. In: 2019
  IEEE Symposium on Security and Privacy (SP). pp. 985--999. IEEE (2019)

\bibitem{chen2016shreds}
Chen, Y., Reymondjohnson, S., Sun, Z., Lu, L.: Shreds: Fine-grained execution
  units with private memory. In: 2016 IEEE Symposium on Security and Privacy
  (SP). pp. 56--71. IEEE (2016)

\bibitem{cox2017efficient}
Cox, G., Bhattacharjee, A.: Efficient address translation for architectures
  with multiple page sizes. ACM SIGOPS Operating Systems Review
  \textbf{51}(2),  435--448 (2017)

\bibitem{deng2015iris}
Deng, Z., Saltaformaggio, B., Zhang, X., Xu, D.: iris: Vetting private api
  abuse in ios applications. In: Proceedings of the 22nd ACM SIGSAC Conference
  on Computer and Communications Security. pp. 44--56. ACM (2015)

\bibitem{durumeric2014matter}
Durumeric, Z., Li, F., Kasten, J., Amann, J., Beekman, J., Payer, M., Weaver,
  N., Adrian, D., Paxson, V., Bailey, M., et~al.: The matter of heartbleed. In:
  Proceedings of the 2014 conference on internet measurement conference. pp.
  475--488. ACM (2014)

\bibitem{elliott2018checked}
Elliott, A.S., Ruef, A., Hicks, M., Tarditi, D.: Checked c: Making c safe by
  extension. In: 2018 IEEE Cybersecurity Development (SecDev). pp. 53--60. IEEE
  (2018)

\bibitem{ferraiuolo2018hyperflow}
Ferraiuolo, A., Zhao, M., Myers, A.C., Suh, G.E.: Hyperflow: A processor
  architecture for nonmalleable, timing-safe information flow security. In:
  Proceedings of the 2018 ACM SIGSAC Conference on Computer and Communications
  Security. pp. 1583--1600. ACM (2018)

\bibitem{frassetto2017jitguard}
Frassetto, T., Gens, D., Liebchen, C., Sadeghi, A.R.: Jitguard: hardening
  just-in-time compilers with {SGX}. In: Proceedings of the 2017 ACM SIGSAC
  Conference on Computer and Communications Security. pp. 2405--2419. ACM
  (2017)

\bibitem{gruss2017kaslr}
Gruss, D., Lipp, M., Schwarz, M., Fellner, R., Maurice, C., Mangard, S.: Kaslr
  is dead: long live kaslr. In: International Symposium on Engineering Secure
  Software and Systems. pp. 161--176. Springer (2017)

\bibitem{guan2017trustshadow}
Guan, L., Liu, P., Xing, X., Ge, X., Zhang, S., Yu, M., Jaeger, T.:
  Trustshadow: Secure execution of unmodified applications with {ARM}
  {TrustZone}. In: Proceedings of the 15th Annual International Conference on
  Mobile Systems, Applications, and Services. pp. 488--501. ACM (2017)

\bibitem{hsu2016enforcing}
Hsu, T.C.H., Hoffman, K., Eugster, P., Payer, M.: Enforcing least privilege
  memory views for multithreaded applications. In: Proceedings of the 2016 ACM
  SIGSAC Conference on Computer and Communications Security. pp. 393--405. ACM
  (2016)

\bibitem{hunt2019isolation}
Hunt, T., Jia, Z., Miller, V., Rossbach, C.J., Witchel, E.: Isolation and
  beyond: Challenges for system security. In: Proceedings of the Workshop on
  Hot Topics in Operating Systems. pp. 96--104. ACM (2019)

\bibitem{mpk}
Intel: Intel® 64 and ia-32 architectures software developer’s manual (2019),
  \url{https://software.intel.com/sites/default/files/managed/39/c5/325462-sdm-vol-1-2abcd-3abcd.pdf}

\bibitem{kocher2018spectre}
Kocher, P., Genkin, D., Gruss, D., Haas, W., Hamburg, M., Lipp, M., Mangard,
  S., Prescher, T., Schwarz, M., Yarom, Y.: Spectre attacks: Exploiting
  speculative execution. arXiv preprint arXiv:1801.01203  (2018)

\bibitem{koning2017no}
Koning, K., Chen, X., Bos, H., Giuffrida, C., Athanasopoulos, E.: No need to
  hide: Protecting safe regions on commodity hardware. In: Proceedings of the
  Twelfth European Conference on Computer Systems. pp. 437--452. ACM (2017)

\bibitem{krohn2007information}
Krohn, M., Yip, A., Brodsky, M., Cliffer, N., Kaashoek, M.F., Kohler, E.,
  Morris, R.: Information flow control for standard os abstractions. In: ACM
  SIGOPS Operating Systems Review. vol.~41, pp. 321--334. ACM (2007)

\bibitem{lamowski2017sandcrust}
Lamowski, B., Weinhold, C., Lackorzynski, A., H{\"a}rtig, H.: Sandcrust:
  Automatic sandboxing of unsafe components in {Rust}. In: Proceedings of the
  9th Workshop on Programming Languages and Operating Systems. pp. 51--57. ACM
  (2017)

\bibitem{lazar2014does}
Lazar, D., Chen, H., Wang, X., Zeldovich, N.: Why does cryptographic software
  fail? a case study and open problems. In: Proceedings of 5th Asia-Pacific
  Workshop on Systems. pp.~1--7 (2014)

\bibitem{lipp2018meltdown}
Lipp, M., Schwarz, M., Gruss, D., Prescher, T., Haas, W., Mangard, S., Kocher,
  P., Genkin, D., Yarom, Y., Hamburg, M.: Meltdown. arXiv preprint
  arXiv:1801.01207  (2018)

\bibitem{litton2016light}
Litton, J., Vahldiek-Oberwagner, A., Elnikety, E., Garg, D., Bhattacharjee, B.,
  Druschel, P.: Light-weight contexts: An {OS} abstraction for safety and
  performance. In: 12th {USENIX} Symposium on Operating Systems Design and
  Implementation ({OSDI} 16). pp. 49--64 (2016)

\bibitem{melara2019enclavedom}
Melara, M.S., Freedman, M.J., Bowman, M.: Enclavedom: Privilege separation for
  large-tcb applications in trusted execution environments. arXiv preprint
  arXiv:1907.13245  (2019)

\bibitem{mo2022sok}
Mo, F., Tarkhani, Z., Haddadi, H.: Sok: Machine learning with confidential
  computing. arXiv preprint arXiv:2208.10134  (2022)

\bibitem{breach2}
Morgan, L.: List of data breaches and cyber attacks in {October} 2017 – 55
  million records leaked.
  \url{https://www.itgovernance.co.uk/blog/list-of-data-breaches-and-cyber-attacks-in-october-2017-55-million-records-leaked/}
  (2017)

\bibitem{morris2002linux}
Morris, J., Smalley, S., Kroah-Hartman, G.: Linux security modules: General
  security support for the linux kernel. In: USENIX Security Symposium. pp.
  17--31. ACM Berkeley, CA (2002)

\bibitem{park2018libmpk}
Park, S., Lee, S., Xu, W., Moon, H., Kim, T.: libmpk: Software abstraction for
  intel memory protection keys. arXiv preprint arXiv:1811.07276  (2018)

\bibitem{provos2003preventing}
Provos, N., Friedl, M., Honeyman, P.: Preventing privilege escalation. In:
  USENIX Security Symposium (2003)

\bibitem{roy2009laminar}
Roy, I., Porter, D.E., Bond, M.D., McKinley, K.S., Witchel, E.: Laminar:
  Practical fine-grained decentralized information flow control, vol.~44. ACM
  (2009)

\bibitem{sehr2010adapting}
Sehr, D., Muth, R., Biffle, C.L., Khimenko, V., Pasko, E., Yee, B., Schimpf,
  K., Chen, B.: Adapting software fault isolation to contemporary cpu
  architectures  (2010)

\bibitem{sigurbjarnarson2018nickel}
Sigurbjarnarson, H., Nelson, L., Castro-Karney, B., Bornholt, J., Torlak, E.,
  Wang, X.: Nickel: a framework for design and verification of information flow
  control systems. In: 13th {USENIX} Symposium on Operating Systems Design and
  Implementation ({OSDI} 18). pp. 287--305 (2018)

\bibitem{singh2021enclaves}
Singh, J., Cobbe, J., Quoc, D.L., Tarkhani, Z.: Enclaves in the clouds: Legal
  considerations and broader implications. Communications of the ACM
  \textbf{64}(5),  42--51 (2021)

\bibitem{iotall}
StewardJack, J.: The ultimate list of internet of things statistics for 2022.
  \url{https://findstack.com/internet-of-things-statistics/} (2021)

\bibitem{tarkhani2022secure}
Tarkhani, Z.: Secure Programming with Dispersed Compartments. Ph.D. thesis,
  University of Cambridge (2022)

\bibitem{tarkhani2020enclave}
Tarkhani, Z., Madhavapeddy, A.: Enclave-aware compartmentalization and secure
  sharing with sirius. arXiv preprint arXiv:2009.01869  (2020)

\bibitem{tarkhani2019snape}
Tarkhani, Z., Madhavapeddy, A., Mortier, R.: Snape: The dark art of handling
  heterogeneous enclaves. In: Proceedings of the 2nd International Workshop on
  Edge Systems, Analytics and Networking. pp. 48--53 (2019)

\bibitem{tarkhani2022enhancing}
Tarkhani, Z., Qendro, L., Brown, M.O., Hill, O., Mascolo, C., Madhavapeddy, A.:
  Enhancing the security \& privacy of wearable brain-computer interfaces.
  arXiv preprint arXiv:2201.07711  (2022)

\bibitem{mpu}
tock: Finer grained memory protection on cortex-m3 mpus.
  \url{https://github.com/tock/tock/issues/1532}

\bibitem{vahldiek2019erim}
Vahldiek-Oberwagner, A., Elnikety, E., Duarte, N.O., Sammler, M., Druschel, P.,
  Garg, D.: {ERIM}: Secure, efficient in-process isolation with protection keys
  ({MPK}). In: 28th {USENIX} Security Symposium ({USENIX} Security 19). pp.
  1221--1238 (2019)

\bibitem{wahbe1994efficient}
Wahbe, R., Lucco, S., Anderson, T.E., Graham, S.L.: Efficient software-based
  fault isolation. In: ACM SIGOPS Operating Systems Review. vol.~27, pp.
  203--216. ACM (1994)

\bibitem{wang2015between}
Wang, J., Xiong, X., Liu, P.: Between mutual trust and mutual distrust:
  practical fine-grained privilege separation in multithreaded applications.
  In: 2015 {USENIX} Annual Technical Conference ({USENIX} {ATC} 15). pp.
  361--373 (2015)

\bibitem{watson2015cheri}
Watson, R.N., Laurie, B., Murdoch, S.J., Norton, R., Roe, M., Son, S., Vadera,
  M., Woodruff, J., Neumann, P.G., Moore, S.W., et~al.: Cheri: A hybrid
  capability-system architecture for scalable software compartmentalization.
  In: 2015 IEEE Symposium on Security and Privacy (SP). pp. 20--37. IEEE (2015)

\bibitem{zeldovich2006making}
Zeldovich, N., Boyd-Wickizer, S., Kohler, E., Mazi{\`e}res, D.: Making
  information flow explicit in histar. In: Proceedings of the 7th symposium on
  Operating systems design and implementation. pp. 263--278. USENIX Association
  (2006)

\bibitem{zeldovich2008hardware}
Zeldovich, N., Kannan, H., Dalton, M., Kozyrakis, C.: Hardware enforcement of
  application security policies using tagged memory. In: OSDI. vol.~8, pp.
  225--240 (2008)

\bibitem{tlbbug}
Zero, P.: Introduction: Bugs in memory management code (2019),
  \url{https://googleprojectzero.blogspot.com/2019/01/taking-page-from-kernels-book-tlb-issue.html}

\bibitem{zhou2014armlock}
Zhou, Y., Wang, X., Chen, Y., Wang, Z.: Armlock: Hardware-based fault isolation
  for {ARM}. In: Proceedings of the 2014 ACM SIGSAC conference on computer and
  communications security. pp. 558--569. ACM (2014)

\end{thebibliography}
